\documentclass{article}
	\usepackage{fullpage, graphicx}
	\author{Sujit S. Datta$^{1}$, Douglas R. Strachan$^{1, 2}$, E. J. Mele$^{1}$, A. T. Charlie Johnson$^{1}$\thanks{E-mail: \texttt{cjohnson@physics.upenn.edu}}\\ \\ {\normalsize 1 -- Department of Physics and Astronomy,}\\ {\normalsize 2 -- Department of Materials Science and Engineering},\\{\normalsize University of Pennsylvania, Philadelphia PA 19104}}	
	\date{{\normalsize Revised Manuscript Submitted: June 20, 2008; }}
	\title{Surface Potentials and Layer Charge\\ Distributions in Few-Layer Graphene Films}
	\begin{document}
	\maketitle
\begin{abstract}
Graphene-derived nanomaterials are emerging as ideal candidates for postsilicon electronics.  Elucidating the electronic interaction between an insulating substrate and few-layer graphene (FLG) films is crucial for device applications.  Here we report electrostatic force microscopy (EFM) measurements revealing that the FLG surface potential increases with film thickness, approaching a ``bulk'' value for samples with five or more graphene layers. This behavior is in sharp contrast with that expected for conventional conducting or semiconducting films, and derives from unique aspects of charge screening by grapheneÕs relativistic low energy carriers. EFM measurements resolve previously unseen electronic perturbations extended along crystallographic directions of structurally disordered FLGs, likely resulting from long-range atomic defects. These results have important implications for graphene nanoelectronics and provide a powerful framework by which key properties can be further investigated.  \end{abstract}
\begin{center}
\line(1,0){300}\\ 
\end{center}
Graphene-derived nanomaterials are a promising family of structures for application as atomically thin transistors, sensors, and other nanoelectronic devices.$^{1-3}$  Graphene, a honeycomb sheet of $sp^{2}$-bonded carbon atoms, and nanotubes, graphene sheets rolled into seamless molecular cylinders, share a set of remarkable electronic properties making them ideal for use in nanoelectronics: tunable carrier type and density, exceptionally high carrier mobility, and structural control of their electronic band structures.$^{4-6}$  A significant advantage of graphene is its two-dimensionality, making it compatible with existing planar device architectures.  This application requires fabrication processes where few-layer graphene (FLG) films are supported on insulating substrates.  While prior theoretical work considered the effect of a substrate on the electronic structure of FLG,$^{7-12}$ few experiments have directly probed the graphene-substrate interaction.$^{10, 13}$  Quantitative understanding of charge exchange at the interface and the spatial distribution of the resulting charge carriers is a critical input to device design. \\

Here we use electrostatic force microscopy (EFM) to probe the electrostatic interactions within FLG samples on SiO$_{2}$ substrates.  We observe the effects of charge exchange between the FLG and the substrate on the surface potential, and we use its variation with thickness to quantify the layer charge distribution in the FLG.  For films less than five layers thick, we observe substantial reduction in the electrostatic potential measured at the exterior graphite surface as the number of layers decreases.  We infer that the charge distribution induced in the FLG differs significantly from that expected for the conventional situation of a conducting or semiconducting thin film. We propose a nonlinear Thomas-Fermi theory for the FLG charge carriers and find excellent quantitative agreement with the data.  The theory predicts that as the surface potential asymptotically approaches the thick film value, it is reduced by a universal finite thickness correction; the analytic form of this correction and its magnitude are in excellent accord with our measurements.  These results highlight the importance of charge exchange between the FLG and the substrate, and provide a general framework for interpreting such interactions in this important family of vertically integrated structures. \\

Our samples consist of FLG sheets transferred onto 300 nm SiO$_{2}$/Si by mechanical exfoliation under ambient.$^{14}$  A Veeco Dimension 3100 atomic force microscope (AFM) with a conductive tip is used to record the sample topography and EFM phase in a two-pass mode.  In the first pass, the AFM tip traces the sample topography (figures 1 and 2A).  In the second (interleave) pass along the same scan line, the tip is biased with a DC voltage $V_{tip}$ and retraces the topography at a fixed lift height $h$ above the surface.  The cantilever is mechanically driven on resonance, and the phase shift of the cantilever oscillation is measured as a function of tip position. Modeling the cantilever as a harmonic oscillator of resonant frequency $\omega_{0}$, spring constant $k$, and quality factor $Q$, and adopting the standard convention that the measured phase shift $\Phi = \phi + \pi/2$ (where $\phi$ is the phase shift between the driving force and the cantilever oscillation), the phase shift over the sample, due to tip-sample capacitive coupling, is:$^{15, 16}$
\begin{eqnarray}
\Phi(x,y)=-\frac{Q}{2k}\Delta C''(h)\left(V_{tip}-V_{s}(x,y)\right)^{2}
\end{eqnarray}
where $C''(h)$ is the second derivative of the difference between the tip-sample capacitance and the tip-bare substrate capacitance as a function of $h$, and $V_{s}(x,y)$ is the local electrostatic potential on the sample surface.  The phase shift of equation (1) is zero when $V_{tip}$ equals the value of $V_{s}$ directly below the tip, so the surface potential can be mapped by EFM with spatial resolution of order 20 nm. EFM is an established technique for mapping the surface potential of planar samples, including semiconductor nanocrystals, thin films of polymers and polymer blends, and self-assembled monolayers.$^{17-21}$\\

From the measured sample topography, we identify well-defined FLG regions with numbers of graphene layers $n$ ranging from 2 to 18.  Figure 2A is a typical topographic image, and figures 2B-C the corresponding EFM phase images for two different tip voltages. The phase shift $\Delta\Phi$ of the FLG regions with respect to the bare SiO$_{2}$ substrate is always negative.  FLG regions with different numbers of graphene layers exhibit different values of $\Delta\Phi$ (figure 2d); fluctuations in $\Delta\Phi$ within each bulk FLG region are indistinguishable from instrumental noise. The relative contrast between FLG regions is reversed in images taken with tip voltages of opposite polarity, indicative of different values of $V_{s}$ (see below).  The data indicate that the surfaces of each FLG region are equipotentials, with surface potential that varies with $n$.\\

We quantify $V_{s}$ by measuring $\Delta\Phi$ as a function of $V_{tip}$, as shown in figure 3A for FLGs that are two and five layers thick.  The maxima of the parabolic fits to the data (see equation (1)) lie at the different values of $V_{s}$.  Figure 3B shows the variation in $\Delta V_{s}=V_{s}-V_{s}^{max}$ as a function of thickness for FLGs with layer number ranging from 2 to 18.  The surface potential is always positive, indicating hole doping in ambient.$^{1}$  For FLGs thicker than 5 layers ($\sim$1.7 nm), the surface potential is approximately constant ($V_{smax}$ = 779 mV), and $V_{s}$ decreases sharply for thinner FLGs.  \\

The central result of our EFM experiments is that the surface potential above FLG films varies strongly (over hundreds of meV) and monotonically with the film thickness.  In the following paragraphs, we model the data quantitatively by assuming that the variation in surface potential is determined by the spatial distribution of doped charges (holes) that are transferred to the graphene from a thin ($<$ 1 nm) interfacial layer of traps or defects at the silica surface. Based on measured current-backgate voltage characteristics of exfoliated FLG conducted in our lab and elsewhere,$^{1-3}$ we expect an areal carrier number density $\sigma_{0}\sim10^{12}$cm$^{-2}$.  With these assumptions, we can use the thickness dependence of $V_{s}$ to probe the layer-by-layer charge distribution in the FLG.  We find that this model of intrinsic screening in the FLG provides an excellent quantitative description of our observations.  This is consistent with the expectation that extrinsic contributions to the surface potential (e.g., charge contamination and adsorbed water) typically lead to a uniform shift in surface potential that does not vary with FLG thickness. \\

The graphene charge distribution is determined by competition between kinetic energy of the doped carriers and their interaction with a self-consistent electrostatic potential.  We describe it with a nonlinear Thomas-Fermi (TF) theory for the doped carriers in the continuum limit, appropriate when the layer areal carrier density $\sigma(z)$ changes smoothly on a scale of the interlayer spacing $d$, assumed to be much less than the FLG thickness $D$.  In the theory, the kinetic energy of the doped carriers in each layer increases with areal number density as $\sigma(z)^{3/2}$ instead of $\sigma(z)^{2}$ as it would for a conventional compressible (conducting) system.  This crucial difference reflects the massless low energy graphene dispersion relation for doped carriers used in the non-linear TF theory, which does not treat the effect of quantum coherence between the graphene layers.  The assumption of decoupled graphene layers is accurate when the local Fermi energies are far from the charge neutrality point and their shifts are large compared to the size of the interlayer hopping amplitudes.  We also note that this was the model successfully applied in references 22 and 23 to study intercalated graphite and has been found$^{24}$ to have good agreement with a self-consistent tight binding model in the range of layer charge densities relevant to our FLG samples.\\

The theory is formulated in terms of a quantity $f(z)=\sqrt{\sigma(z)}$ which is a solution to the nonlinear equation
\begin{eqnarray}
\frac{d^{2}f(z)}{dz^{2}}=\frac{2\tilde{\beta}}{d}f^{2}(z)
\end{eqnarray}
where $\gamma=2\sqrt{\pi}\hbar v_{F}/3$, $\tilde{\beta}=4\pi e^{2}/3\gamma$, subject to the boundary conditions $[df/dz]_{z=0}=2\tilde{\beta}\sigma_{0}$ and $[df/dz]_{z=D}=0$ (see supporting text for details).  To understand the solutions we scale $f(z)$ by its maximum value $f_{0}$ at the interface and define the ratio parameter $r_{D}=f(D)/f_{0}$ by the integral relation
\begin{eqnarray}
(1-r_{D}^{3})^{1/6}\int_{1}^{r_{D}}\frac{du}{\sqrt{u^{3}-r_{D}^{3}}}=2\left(\frac{\tilde{\beta}\sigma_{0}D^{3}}{3d}\right)^{1/3}=\Gamma
\end{eqnarray}
The dimensionless parameter $\Gamma$ determines $r_{D}$ and consequently the nature of the FLG screening.  When $\Gamma>1$ the system is in a strong coupling regime dominated by its electrostatic energy;  here $r_{D}\ll1$ and the doped hole distribution is strongly peaked near the oxide interface.  The FLG contribution to the measured $V_{s}$ can be expressed in terms of $r_{D}$, giving the central result of the Thomas-Fermi theory
\begin{eqnarray}
V_{s}=\frac{3\gamma}{2}(3\tilde{\beta}d\sigma_{0}^{2})^{1/3}\frac{1-r_{D}}{(1-r_{D}^{3})^{1/3}}+V_{0}
\end{eqnarray}
where $V_{0}=3\gamma\tilde{\beta}\sigma_{0}d_{s}/2$ is a constant offset due to the charge distribution in the substrate. Here $d_{s}$ is the characteristic depth of ionized acceptors in the SiO$_{2}$, estimated to be $0.5-1$nm if the charge resides in an interfacial defect or impurity layer, as expected.  Thus the graphene contribution dominates the surface potential whenever $\sigma_{0}<2d/(\tilde{\beta}^{2}d_{s}^{3})\sim6\times10^{12}$cm$^{-2}$.  In this situation, variations in the surface potential $V_{s}$ with film thickness $D$ are controlled entirely by the distribution of charge in the FLG in the experimentally relevant situation where $\sigma_{0}$ is independent of $D$.$^{13}$\\

Our samples are described by the strong coupling limit of the TF model, for which we find $V_{s}\approx (3\gamma/2)(3\tilde{\beta}d\sigma_{0}^{2})^{1/3}-9\gamma d/2\tilde{\beta}D^{2}$.  This explains the observed asymptotic behavior of $V_{s}$, with the thick film value ($V_{s}^{max}$) determined by the areal charge density. As it approaches this limit, $V_{s}$ contains a universal (independent of $\sigma_{0}$) finite size ``thin film" correction. Using the accepted values $\gamma=6.5$eV$\cdot\AA$ and $\tilde{\beta}=9.5$, we calculate the finite size correction 
\begin{eqnarray}
V_{s}^{max}-V_{s}=\Delta V_{s}=-10.3\kappa(\mbox{V}\cdot\AA^{2})/D^{2}
\end{eqnarray}
including the effect of a static dielectric constant $\kappa$ perpendicular to the layers. Although this $1/D^{2}$ dependence resembles the confinement energy of a quantum mechanical gas of massive doped holes, here it arises from a competition between the two-dimensional kinetic energy for a gas of massless carriers (which suppresses spatial variations in $\sigma(z)$), and the Coulomb interactions in the third dimension (which favors them). Appropriate values of the dielectric constant for the FLG/SiO$_{2}$ interface region lie in the range 2 $<$ $\kappa$ $<$ 3;$^{22-24}$  the solid lines in figure 3 correspond to these lower and upper bounds, while the dotted line (using $\kappa$ = 2.5) gives an excellent fit to the data.  Though our present measurements do not exclude the possibility of an exponential decay (as expected for screening by massive carriers), the excellent quantitative agreement with eq 5 gives strong support for the non-linear Thomas-Fermi behavior along the $c$-axis in FLG samples. 
\\

Solving equation (2) with $d_{s}=0.5$nm and $V_{s}^{max}=779$mV gives $\sigma_{0}=6\times10^{12}$cm$^{-2}$, which is understood as arising from charge exchange to a small acceptor density at the interface.  This indicates that for our samples, the potential drop in the substrate and the FLG contribute roughly equally to the overall surface potential.  Furthermore, it confirms that our FLG films realize the strong coupling limit of the model ($\Gamma>1$) where the doped holes accumulate close to the interface with the oxide, with a strong power law decay into the FLG film. The rapid nature of this decay can be inferred directly from the rapid convergence of the surface potential $V_{s}$ to its thick film limit (figure 3B).\\

Our data is consistent with recent angle-resolved photoemission spectroscopy measurements of FLG on SiC;$^{13}$ indeed, neither study finds any evidence for the oscillatory charge and potential profile that are predicted from a linear response theory of the screening.$^{25}$  It is interesting to note that our analysis predicts that the induced charge profile in the graphene film has a power law decay, a situation that is thought to describe the layer charge distributions found in graphite intercalation compounds.$^{22-24}$  Furthermore, our model of charge exchange to a small density of defect/interface states is quite similar to conclusions reached via qualitative considerations of Fermi energy equilibration for pentacene films on SiO$_{2}$.$^{26}$ \\

These results suggest that EFM might also be used to investigate electronic perturbations in FLG films.  In particular, structural disorder is expected to strongly alter FLG electronic properties$^{27, 28}$ in ways yet to be understood.  To this end, we extend EFM measurements to the study of disordered FLGs, specifically a representative sample consisting of 9 graphene layers in the bulk (figure 4A).  The presence of well-defined folds suggests a high density of defects and dislocations in the FLG, and associated local stresses within it.$^{29, 30}$  A comparison to the schematic in figure 4C shows that the many folds along the right edge of the film all fall along $\left<100\right>$ ``zig-zag" and $\left<100\right>$ ``armchair" directions of the graphene lattice. \\

Although mechanical stresses and folding along crystallographic directions of graphitic samples have been studied,$^{29, 30}$ they have not been correlated with electronic properties.  Strikingly, EFM phase images of the disordered FLG film indicate parallel stripe-like electronic perturbations along specific directions of the underlying honeycomb lattice (figure 4B). Topographical features corresponding to these stripes are not discernable by AFM. For clarity, the ends of one such stripe are indicated in figure 4B with arrows, where the potential over the stripe is reduced by $\sim$0.25V compared to the unperturbed bulk region, where we find $V_{s}$ = 0.76V, in agreement with figure 3.  Remarkably, all observed electronic stripe perturbations lie parallel ($\pm3^\circ$) to a single fold direction of the underlying graphene lattice.  Although we do not know the precise origin of these electronic perturbations, there are several possible sources, including grain boundaries or microcracks,$^{26}$ `scars',$^{25}$ or $sp^{3}$-like rehybridization defects.$^{29, 30}$ Interestingly, similar electronic perturbations were observed at surface defects such as line dislocations and grain boundaries in pentacene multilayers.$^{26}$ The data suggest the intriguing possibility that EFM could prove useful for nanoscale imaging of local defects and their impact on electronic structure in graphene-based devices. \\
\newpage
\begin{center}\noindent{\bf Acknowledgment}\end{center}
We thank Y. Zhao (P. Kim group, Columbia University) for help with sample preparation. Research supported by the Nano/Bio Interface Center through the National Science Foundation NSEC DMR-0425780; the JSTO DTRA, the Army Research Office Grant W911NF-06-1-0462; the Department of Energy Grant DE-FG02-06ER45118; and the Intelligence Community Postdoctoral Fellowship Program.\\ \\
\begin{center}\noindent{\bf References}\end{center}
\begin{enumerate}
\item Novoselov, K. S. {\it et al}. Electric Field Effect in Atomically Thin Carbon Films. {\it Science} {\bf 306,} 666 - 669 (2004).
\item Schedin, F. {\it et al}. Detection of individual gas molecules adsorbed on graphene. {\it Nature Materials}, Advanced Online Publication (2007).
\item Williams, J. R., DiCarlo, L., Marcus, C. M. Quantum Hall Effect in a Gate-Controlled p-n Junction of Graphene. {\it Science} {\bf 317,} 638-341 (2007).
\item Novoselov, K. S. {\it et al}. Two-Dimensional Gas of Massless Dirac Fermions in Graphene. {\it Nature} {\bf 438,} 197 - 200 (2005).
\item Ohta, T., Bostwick, A., Seyller, T., Horn, K., Rotenberg, E. Controlling the Electronic Structure of Bilayer Graphene. {\it Science} {\bf 313,} 951 - 954 (2006).
\item Zhang, Y., Tan, Y.-W., Stormer, H. L., Kim, P. Experimental observation of the quantum Hall effect and Berry's phase in graphene. {\it Nature} {\bf 438,} 201 - 204 (2005).
\item Guinea, F. Charge Distribution and Screening in Layered Graphene Systems. {\it Physical Review B} {\bf 75,} 235433 (2007).
\item McCann, E. Asymmetry gap in the electronic band structure of bilayer graphene. {\it Physical Review B} {\bf 74,} 161403(R) (2006).
\item Min, H., Sahu, B., Banerjee, S. K., MacDonald, A. H. Ab Initio Theory of Gate Induced Gaps in Graphene Bilayers. {\it Physical Review B} {\bf 75,} 155115 (2007).
\item Zhou, S. Y. {\it et al}. Substrate-induced bandgap opening in epitaxial graphene. {\it Nature Materials} {\bf 6,} 770-775 (2007).
\item Adam, S., Hwang, E. H., Galitski, V. M., Sarma, S. D. A self-consistent theory for graphene transport. {\it Proc. Natl. Acad. Sci.} {\bf 104,} 18392 (2007).
\item Katsnelson, M. I., Geim, A. K. Electron scattering on microscopic corrugations in graphene. {\it Phil. Trans. R. Soc. A} {\bf 366,} 195 (2008).
\item Ohta, T. {\it et al}. Interlayer Interaction and Electronic Screening in Multilayer Graphene Investigated with Angle-Resolved Photoemission Spectroscopy. {\it Physical Review Letters} {\bf 98,} 206802 (2007).
\item Information on materials and methods is available online.
\item Staii, C., Johnson, A. T., Pinto, N. J. Quantitative Analysis of Scanning Conductance Microscopy. {\it Nano Letters} {\bf 4,} 859 - 862 (2004).
\item Bockrath, M. {\it et al}. Scanned Conductance Microscopy of Carbon Nanotubes and $\lambda$-DNA. {\it Nano Letters} {\bf 2,} 187-190 (2002).
\item Coffey, D. C., Ginger, D. S. Time-Resolved Electrostatic Force Microscopy of Polymer Solar Cells. {\it Nature Materials} {\bf 5,} 735 - 740 (2006).
\item Lei, C. H., Das, A., Elliott, M., Macdonald, J. E. Quantitative electrostatic force microscopy phase measurements. {\it Nanotechnology} {\bf 15,} 627 (2004).
\item Silveira, W. R., Muller, E. M., Ng, T. N., Dunlap, D. H., Marohn, J. A. in Scanning Probe Microscopy: Electrical and Electromechanical Phenomena at the Nanoscale (eds. Kalinin, S. V., Gruverman, A.) (Springer Verlag, New York, 2005).
\item Takano, H., Wong, S.-S., Harnisch, J. A., Porter, M. D. Mapping the subsurface composition of organic films by electric force microscopy. {\it Langmuir} {\bf 16,} 5231 (2000).
\item Gordon, J. M., Baron, T. Amplitude-mode electrostatic force microscopy in UHV: Quantification of nanocrystal charge storage. {\it Physical Review B} {\bf 72,} 165420 (2005).
\item Pietronero, L., StrŠssler, S., Zeller, H. R., Rice, M. J. Charge Distribution in $c$ Direction in Lamellar Graphite Acceptor Intercalation Compounds {\it Physical Review Letters} {\bf 41,} 763-767 (1978).
\item Safran, S. A., Hamann, D. R. Electrostatic interactions and staging in graphite intercalation compounds. {\it Physical Review B} {\bf 22,} 606-612 (1980).
\item Safran, S. A., Hamann, D. R. Self-consistent charge densities, band structures, and staging energies of graphite intercalation compounds. {\it Physical Review B} {\bf 23,} 565-574 (1981).
\item Guinea, F. Charge distribution and screening in layered graphene systems. {\it Physical Review B} {\bf 75,} 235433 (2007).
\item Puntambekar, K., Dong, J., Haugstad, G., Frisbie, C. D. Structural and Electrostatic Complexity at a Pentacene/Insulator Interface. {\it Advanced Functional Materials} {\bf 16,} 879 (2006).
\item Morozov, S. V. {\it et al}. Strong suppression of weak localization in graphene. {\it Physical Review Letters} {\bf 97,} 016801 (2006).
\item Peres, N. M. R., Guinea, F., Neto, A. H. C. Electronic Properties of Disordered Two-Dimensional Carbon. {\it Physical Review B} {\bf 73,} 125411 (2006).
\item Ebbesen, T. W., Hiura, H. Graphene in 3-Dimensions: Towards Graphite Origami. {\it Advanced Materials} {\bf 7,} 582 (1995).
\item Li, L. X. {\it et al}. Tearing, folding and deformation of a carbon-carbon $sp^{2}$-bonded network. {\it Carbon} {\bf 44,} 1544 (2006).
\end{enumerate}

\newpage

\begin{figure}
\begin{center}
\includegraphics[width=5.5in]{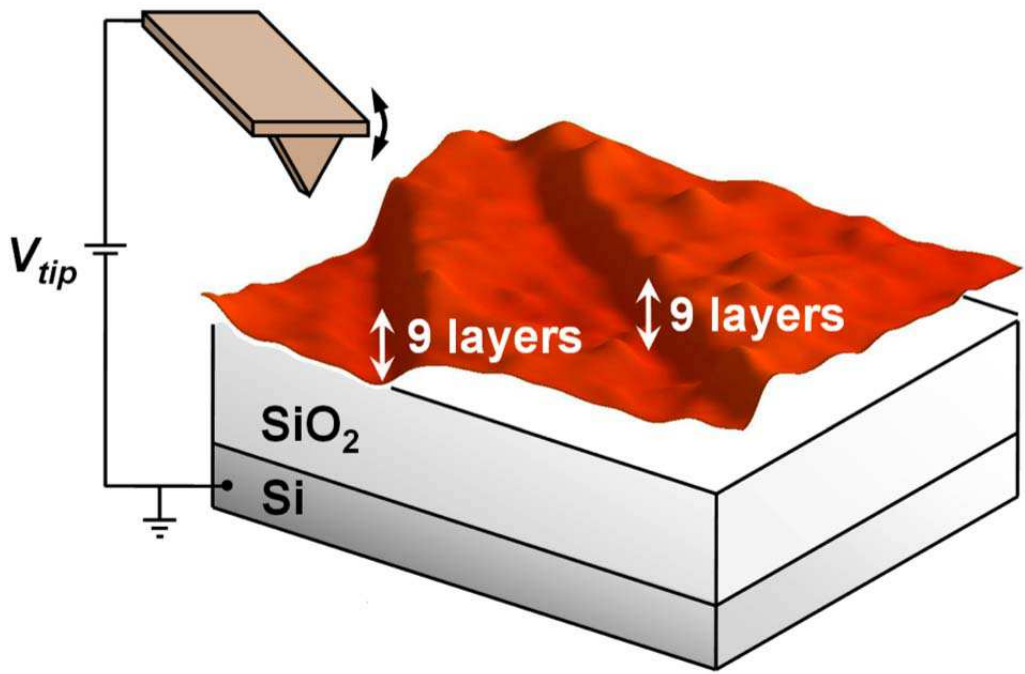}
\caption{Schematic of the experiment. FLG samples of different thicknesses (here, 9 and 18 layers) are deposited on a SiO$_{2}$ substrate on a highly-doped Si ground plane. Sample topography and EFM phase are imaged simultaneously. Orange surface plot is 4.1 $\mu$m $\times$ 4.1 $\mu$m AFM topography image, with low-pass filtering. }
\end{center}
\end{figure}
\newpage

\begin{figure}
\begin{center}
\includegraphics[width=5in]{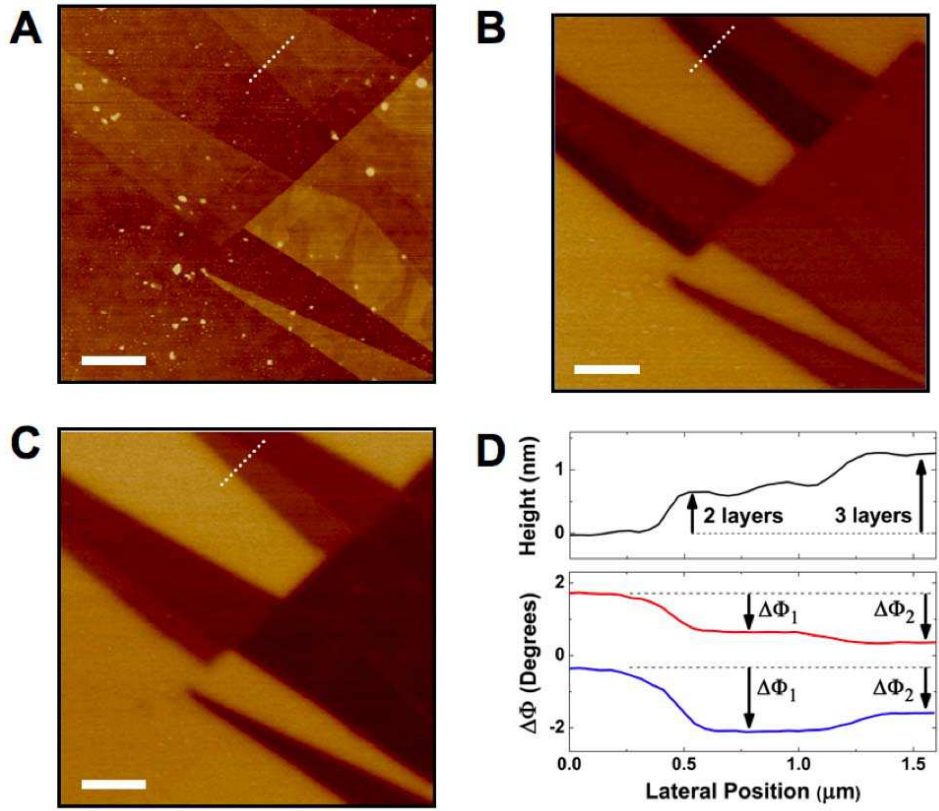}
\caption{AFM and EFM of few-layer graphene. (A) AFM image of FLGs on SiO$_{2}$/Si substrate; color scale is 10nm. (B, C) EFM phase images of the sample, with $V_{tip}$ = -2V and +3V, respectively. Color scales are 5.0$^\circ$¡. (D) Average of 30 line scans of topography and phase centered along the dashed lines in AFM and EFM images, with low-pass filtering. Black curve corresponds to (A), red curve to (B), and blue curve to (C). Scale bar in each image is 1.5 $\mu$m.}
\end{center}
\end{figure}

\newpage

\begin{figure}
\begin{center}
\includegraphics[width=5.5in]{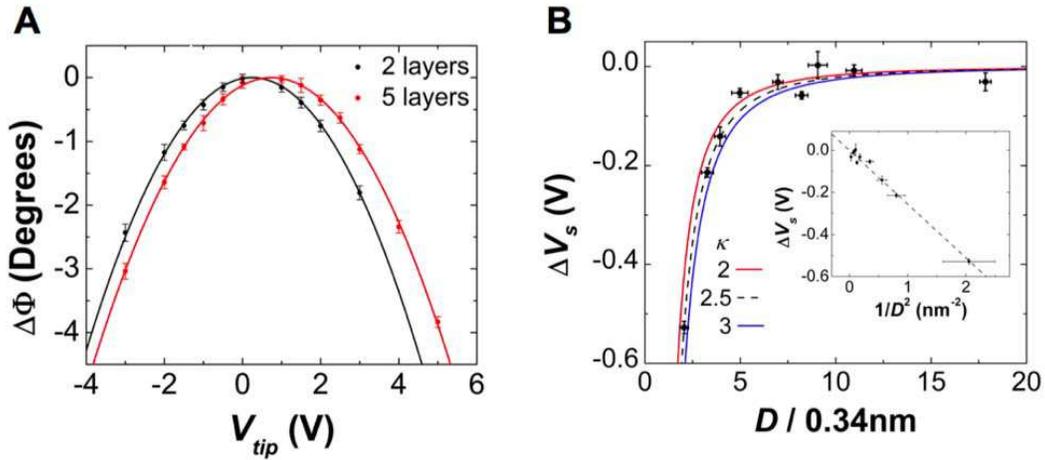}
\caption{Surface potential measurement by EFM. (A) EFM phase shift versus tip voltage for FLGs with 2 and 5 layers (black and red data, respectively). Solid lines are parabolic fits to the data. Error bars represent instrumental noise. (B) Plot of the variation in surface potential, $\Delta V_{s}=V_{s}-V_{s}^{max}$, versus FLG thickness $D$, normalized by single graphene layer thickness (0.34nm). Data points for $D/0.34$nm $\approx 2$, 3, 7, 8 are averages of data from two different regions of equal $D$, while other data points are over a single region of the specified thickness. Error bars represent statistical uncertainty in the value of $V_{s}$ obtained from parabolic fits as in Fig 3A. Curves represent predictions of a nonlinear Thomas-Fermi screening model with background dielectric constant $\kappa$ = 2 (red), 2.5 (dashed), and 3 (blue). Inset: $\Delta V_{s}$ as a function of $1/D^{2}$, with the power-law fit predicted by the model.}
\end{center}
\end{figure}
\newpage

\begin{figure}
\begin{center}
\includegraphics[width=5in]{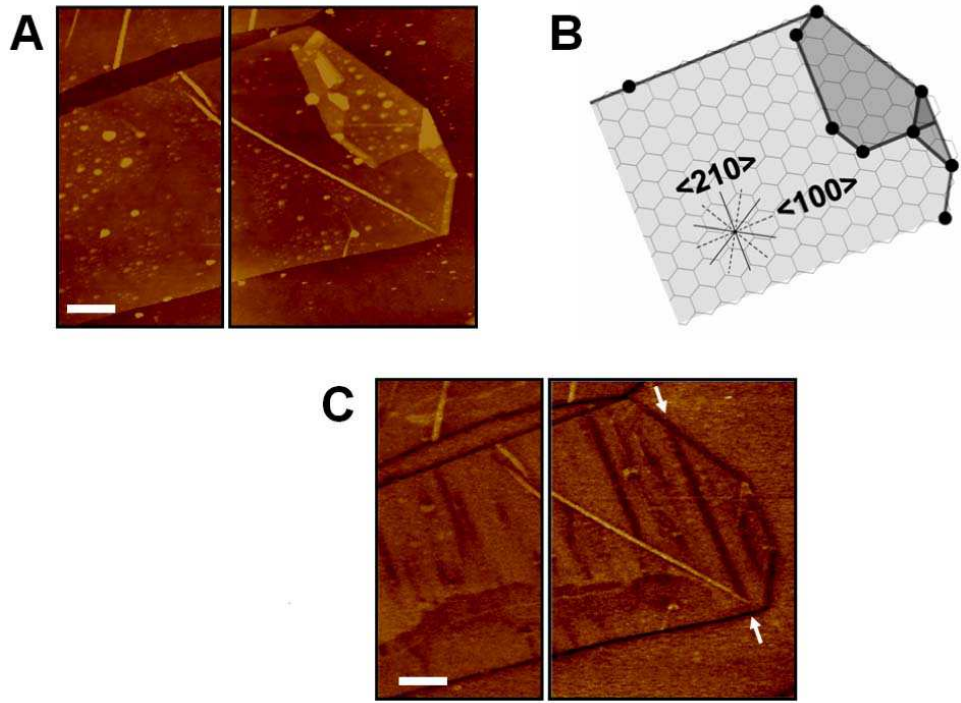}
\caption{Extended defects in few-layer graphene are revealed in EFM. (A) AFM topographic image of a disordered FLG sample; color scale is 10nm. (B) Schematic of a possible structure, indicating folding, symmetry axes, and FLG crystallographic axes. (C) Corresponding EFM phase image, with $V_{tip}$ = +1.5V. Color scale is 1.0$^\circ$. All scale bars are 1.5 $\mu$m.}
\end{center}
\end{figure}

	\end{document}


\maketitle
\begin{center}
\line(1,0){300}\\ 
\end{center}
\section{Materials and Methods}
\subsection{Graphene Preparation}
Graphene samples are synthesized using a mechanical exfoliation technique similar to that described in Ref. 1. We start with a highly-doped Si substrate with 300nm of thermally-grown SiO$_{2}$, cleaned with acetone and isopropyl alcohol. Adhesive scotch tape is used to extract and exfoliate a starting piece of bulk Kish graphite (Toshiba Ceramics, San Jose, CA). After the graphite has been sufficiently thinned, the tape is pressed against the SiO$_{2}$ surface and gently rubbed with the back of a tweezer for approximately 15s. The strength with which the tape is rubbed is varied across the sample surface, resulting in distinct FLG regions, some structurally pristine and others disordered. FLG regions of varying thickness are selected by visual inspection with an optical microscope.\\

\subsection{Details of AFM and EFM measurements}
Two different kinds of metal-coated SPM tips are used for AFM and EFM measurements: Ti-Pt tips (NSC18, Mikromasch) are characterized by force constant $k\sim $ 3.5N/m, quality factor $Q\sim$ 250, resonant frequency $\omega_{0}\sim$ 75kHz, tip curvature radius $\sim$ 40nm; Cr-Au tips (NSC15, Mikromasch) are characterized by $k\sim$ 40N/m, quality factor $Q\sim$ 200, resonant frequency $\omega_{0}\sim$ 325kHz, tip curvature radius $\sim$ 50nm. AFM height measurements are done using intermittent contact mode.\\

\subsection{Identification of Number of Layers}
Previous experiments have indicated the possible presence of a ``dead layer" between the FLG and the SiO$_{2}$ substrate that causes AFM measurements of the graphene thickness to be increased by up to several angstroms$^{1}$. If a dead layer were present, accurate identification of the number of layers in the FLG would be complicated. However, the number of layers present in a FLG region can be accurately measured by AFM if the film contains folds or wrinkles. AFM measurements of a folded FLG region prepared on the same chip as our samples indicate that the thickness of the dead layer, if there is one, is much smaller than the thickness of a single graphene layer (0.34 nm), so we are able to accurately determine the number of graphene layers in our FLG films. An example of this is shown in fig. S1. \\

\subsection{EFM Measurements}
By convention, the EFM phase shift $\Delta\Phi$ is measured with respect to the bare substrate (i.e. $\Delta\Phi= 0$ over a SiO$_{2}$/Si region with no deposited FLG). Each data point of fig. 3A (main text) represents the average of $\sim$5-10 different line scans over the same region (or the average of 10-20 line scans taken over two different regions of identical thickness -- see the caption of Figure 3). As explained in detail in Ref. 2, conducting samples exhibit a negative phase shift, while insulating samples exhibit a positive phase shift that depends on the sample dielectric constant. For electrically floating conducting samples such as FLGs, changes in $\phi(x,y)$ (given by equation 1 of the main text) result from changes in $C''(h)$, a geometric effect, or changes in the local sample surface electrostatic potential $V_{s}(x,y)$. Because the surface roughness of FLG bulk regions is of order 0.1 -- 0.5 nm ($\sim$0.3\% of the tip radius$^{3}$), $C''(h)$ is constant to an excellent approximation, and changes in $\Delta\Phi$ reflect changes in $V_{s}$.$^{4}$\\

We conducted multiple control experiments to verify the accuracy and reproducibility of the surface potential measurement. Height-dependent EFM measurements (fig. S2A) of $\Delta\Phi$ over the same FLG region at fixed $V_{tip}$ have a power-law form $\Delta\Phi\sim h^{-1.6}$, characterized by an exponent between that expected for a cone-plane geometry ($C''(h)\sim h^{-1}$) and a sphere-plane geometry ($C''(h)\sim h^{-2}$), as seen by others$^{5}$. EFM measurements of the same FLG region were taken using two different lift heights, with the $\Delta\Phi$--$V_{tip}$ data shown in Fig S2B. The measured value of $V_{s}$ is verified to be independent of the lift height $h$, as expected. \\

\section{Supporting Text}
\subsection{Outline of Thomas-Fermi Theory}
The Thomas-Fermi (TF) theory describes the distribution of layer areal number densities $\sigma(z)$ that are in equilibrium with respect to fluctuations $\delta\sigma(z)$. The TF theory does not treat the effect of quantum coherence between the layers, and it applies in the controlled limit that the interlayer tunneling is tuned to zero. It is accurate when the variations of the interlayer potential or equivalently, when the local shift of the Fermi energies are large compared to the size of the interlayer hopping amplitudes. \\

For doped carriers described by the conical dispersion relation $E(k)=\hbar v_{F}k$, an excess areal carrier density contributes to the kinetic energy (per unit area) of each layer $K_{i}=2\sqrt{\pi}\hbar v_{F} \sigma_{i}^{3/2}/3$.  The doped carriers in each layer also interact with a compensating density $-e\sigma_{0}$ in the substrate and with the charges in each layer. In the continuum limit, the grand potential for this system is 
\begin{eqnarray}
\Omega=\Omega_{1}+\Omega_{2}
\end{eqnarray}
\begin{eqnarray}
\Omega_{1}=\int_{0}^{D}\frac{dz}{d}\left(\gamma\sigma(z)^{3/2}-\mu\sigma(z)-2\pi e^{2}z\sigma_{0}\sigma(z)\right) 
\end{eqnarray}
\begin{eqnarray}
\Omega_{2}=-\int_{0}^{D}\pi e^{2}\frac{dz}{d}\frac{dz'}{d}\sigma(z)|z-z'|\sigma(z')
\end{eqnarray}
where $\gamma=2\sqrt{\pi}\hbar v_{F}/3$ and $\mu$ is the chemical potential. Minimizing $\Omega$ and defining $f(z)=\sqrt{\sigma(z)}$ one finds
\begin{eqnarray}
\frac{3\gamma}{2}f(z)=\mu+\beta z\sigma_{0}+\frac{\beta}{d}\int_{0}^{D}dz'|z-z'|\sigma(z')
\end{eqnarray}
and differentiating twice with respect to the observer coordinate $z$ we obtain the equation quoted in the text
\begin{eqnarray}
\frac{d^{2}f}{dz^{2}}=\frac{2\tilde{\beta}}{d}f(z)^{2}
\end{eqnarray}
with $\tilde{\beta}=4\pi e^{2}/3\gamma$. The boundary conditions can be determined by considering the behavior of the first derivative of eq. (4), along with the constraint of charge conservation
\begin{eqnarray}
\left[\frac{df}{dz}\right]_{z=0}=\tilde{\beta}\left(\sigma_{0}-\int_{0}^{D}\frac{dz'}{d}\sigma(z')\right)=2\tilde{\beta}\sigma_{0}
\end{eqnarray}
\begin{eqnarray}
\left[\frac{df}{dz}\right]_{z=D}=\tilde{\beta}\left(\sigma_{0}+\int_{0}^{D}\frac{dz'}{d}\sigma(z')\right)=0
\end{eqnarray}
The solutions of eq. (5) are then obtained by using the conservation law
\begin{eqnarray}
\frac{d}{dz}\left(\frac{1}{2}\left(\frac{df}{dz}\right)^{2}-\frac{2\tilde{\beta}}{3d}f^{3}\right)=0
\end{eqnarray}
and then directly integrating $df/dz$ to find $f(z)$. In terms of a dimensionless variable of integration $u=f(z)/f(0)$ we find
\begin{eqnarray}
(1-r_{D}^{3})^{1/6}\int_{1}^{r_{D}}\frac{du}{\sqrt{u^{3}-r_{D}^{3}}}=2\left(\frac{\tilde{\beta}\sigma_{0}D^{3}}{3d}\right)^{1/3}=\Gamma
\end{eqnarray}
where $r_{D}=f(D)/f(0)$. Equation (9) provides a convenient relation between the dimensionless coupling parameter $\Gamma$ and the screening parameter $r_{D}$.
Once $r_{D}$ is determined (4) can be used to find the graphene contribution to the surface potential
\begin{eqnarray}
V_{D}=\frac{3\gamma}{2}(3\tilde{\beta}d\sigma_{0}^{2})^{1/3}\frac{1-r_{D}}{(1-r_{D}^{3})^{1/3}}+V(0)
\end{eqnarray}
The first term on the right hand side is the contribution to the surface potential from the charge distribution in the graphene. $V(0)$ is the potential at the graphene/silica interface and is offset from the potential deep in the silica because of the charge distribution in the charged acceptor layer. Setting the potential to zero in the silica bulk, we have
\begin{eqnarray}
V(0)=\frac{3\gamma}{2}\tilde{\beta}\sigma_{0}d_{s}=2\pi e^{2}\sigma_{0}d_{s}
\end{eqnarray}
for an areal acceptor density $\sigma_{0}$ within depth $d_{s}$ of the interface. \\

\newpage

\begin{center}\noindent{\bf Supporting References}\end{center}
\begin{enumerate}
\item Novoselov, K. S. {\it et al}., {\it Science} {\bf 306,} 666 - 669 (2004).
\item Staii, C., Johnson, A. T., Pinto, N. J., {\it Nano Letters} {\bf 4,} 859 - 862 (2004).
\item Ishigami, M., Chen, J. H., Cullen, W. G., Fuhrer, M. S., Williams, E. D., {\it Nano Letters} {\bf 7,} 1643 (2007).
\item We note that because our measurements are performed under ambient laboratory conditions, the measured $V_{s}$ is expected to be offset from the value measured under UHV. For example, although graphene is well known to be hydrophobic, we can not rule out the presence of a thin adsorbed water layer on the sample surface, or hydrogen bonded to the Si-OH silanol groups at the SiO$_{2}$. However, our investigations are of robust changes in the FLG surface electrostatic potential, and thus the slight offset introduced in our measurements by such effects does not affect our conclusions. Furthermore, we note that all the measurements presented are robust over a period of multiple weeks.
\item Belaidi, S. {\it et al}., {\it J. Appl. Phys.} {\bf 81,} 1023 (1997). 

\end{enumerate}

\newpage

\begin{figure}
\begin{center}
\includegraphics[width=5.5in]{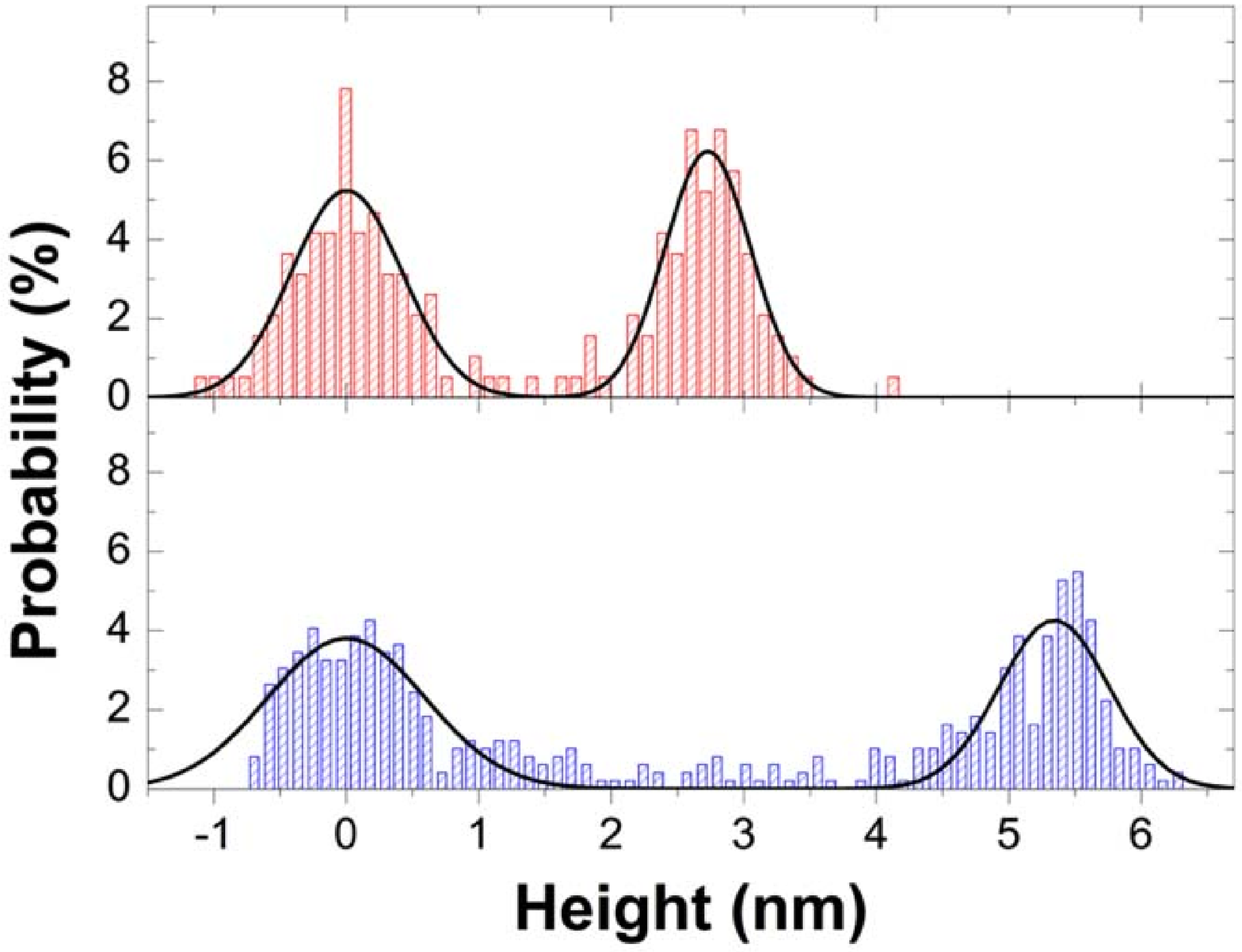}
\caption{AFM across FLG fold. Height histograms acquired across an unfolded FLG/substrate edge (top) and a folded FLG/substrate edge (bottom), for neighboring pristine regions of the sample shown in figure 4 of the main text. The height difference between the bulk region and the SiO$_{2}$ is $\sim$2.73 nm, while the height difference between the folded region and the SiO$_{2}$ is $\sim$5.3 nm, almost exactly twice as large. We thus conclude that the unfolded region consists of 8 layers ($\sim$ 2.73 nm / 0.34 nm, 0.34 nm being the thickness of a single graphene layer), and that the thickness of the ``dead layer" (if any) in this experiment is negligible compared to the interlayer spacing.}
\end{center}
\end{figure}
\newpage

\begin{figure}
\begin{center}
\includegraphics[width=5.5in]{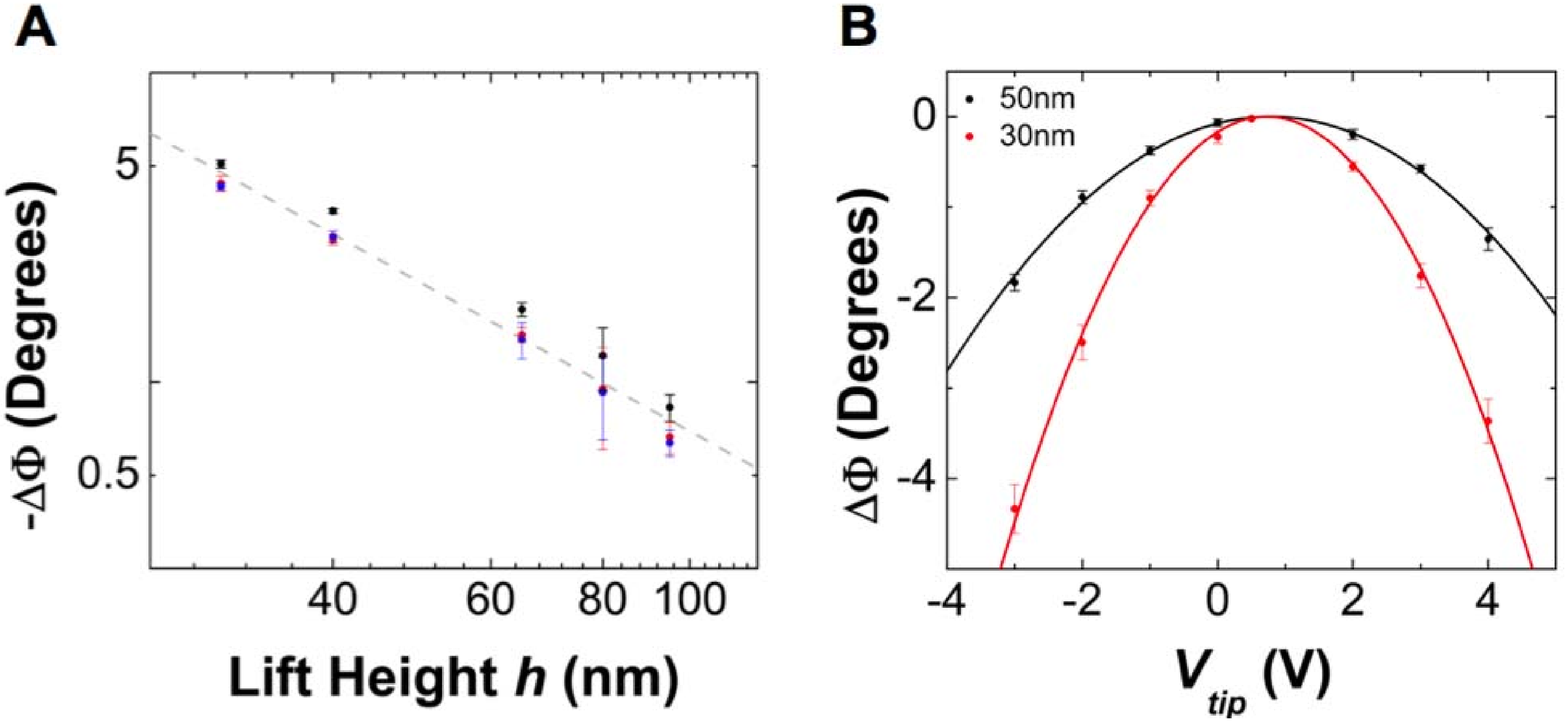}
\caption{Results of EFM control experiments.  (A) Negative of EFM phase shift ($-\Delta\Phi$) measured as a function of the lift height $h$, over three different FLG regions (different colors). The dashed line indicates the functional form $-\Delta\Phi\sim h^{-1.6}$. (B) $\Delta\Phi$ versus tip voltage $V_{tip}$ for the same FLG region of height $\sim$6.1nm (18 layers), for lift heights $h$ = 30nm and 50nm. The measured $\Delta V_{s}=V_{s}-V_{s}^{max}=0.02\pm0.06$V agrees well with data in figure 3 of the main text, and confirms that the measured value of $V_{s}$ is not sensitive to the lift height $h$. }
\end{center}
\end{figure}